\documentclass[a4paper]{jaa}

\usepackage[sort&compress,round,comma,authoryear]{natbib}
\usepackage{latexsym,amsmath,amsfonts,amssymb}
\usepackage{graphicx,epsfig}
\usepackage{color}

\begin{document}
\title{High energy power-law tail in X-ray binaries and bulk Comptonization due to a outflow from a disk}

\author{Nagendra Kumar}
\affilOne{Department of Physics, Indian Institute of Science, Bangalore 560012, India.}


\def\t{\text}
\def\r{\textcolor{red}}
\def\b{\textcolor{blue}}

\twocolumn[{

\maketitle

\corres{nagendra.bhu@gmail.com}


\begin{abstract}
We study the high energy power-law tail emission of X-ray binaries (XRBs) 
by a bulk Comptonization process which usually observe in the very high soft
VHS state of blackhole BH XRBs 
and the high soft HS state of neutron star (NS) and
BH XRBs. 
Earlier, to generate the power-law tail in bulk Comptonization framework,
a free-fall converging flow into BH or NS has been considered as a bulk region.
In this work, for a bulk region we consider mainly an outflow geometry from
the accretion disk which is bounded by a torus surrounding the compact object.
We have a two choice for an outflow geometry i) 
collimated flow  ii) conical flow of
opening angle $\theta_b$ and the axis is perpendicular to the disk.
We also consider an azimuthal velocity of the torus fluids  
as a bulk motion where the fluids are rotating around the compact object
(a torus flow).
We find that the power-law tail can be generated in a torus flow with having
large optical depth and bulk speed ($>$ 0.75c), and in conical flow with
$\theta_b$ $>$ $\sim$30 degree for a low value of Comptonizing medium
temperature.
Particularly, in conical flow the low opening angle is more favourable
to generate the power-law tail in both state HS and VHS.
We notice that when the outflow is collimated, then the
emergent spectrum does not have power-law component for a low Comptonizing medium temperature.

\end{abstract}


}]



\section{Introduction}
X-ray binares (XRBs) generally exhibit three spectral states, a) power-law dominated low intense hard (LH) state, b) blackbody
dominated high intense soft (HS) state and c) intermediate state (IS).
XRBs frequently transit from one spectral state to other spectral state in
sequence, like LH $\rightarrow$ IS $\rightarrow$ HS $\rightarrow$ IS
$\rightarrow$ LH state. In black hole BH XRBs during a spectral transition from LH to HS, a
very high intense power-law dominated (VHS) state is observed, which usually
extends more than 200 keV without an exponential cut-off with
photon index $\Gamma$ $>$ 2.4. VHS state is
also termed as a steep power law (SPL) state. Generally, the SPL state occurs
through an outburst and in the hardness intensity diagram (or Q-diagram), it is
observed in upper-right region, where
occasionally an episodic
jet is observed 
(\citealp[for review, see][]{McClintock-Remillard2006, Done-etal2007, Belloni-etal2011}; \citealp[in particular,][]{Debnath-etal2008, Dunn-etal2010,  Nandi-etal2012, McClintock-etal2009, Pahari-etal2014}).
A high energy power-law tail is also observed in HS state of both BH and NS
(neutron star) XRBs, which may be
extended up to 200 keV or more with $\Gamma$ $>$ 2.0 \cite[see, e.g.,][]{Joinet-etal2005,Motta-etal2009,Titarchuk-Shaposhnikov2010,Revnivtsev-etal2014,Titarchuk-etal2014}. 

The power-law component in X-ray spectrum of XRBs generally explain by thermal
Comptonization process, where the soft photons get upscattered by high
energetic electrons, which have thermal velocity distribution
\cite[e.g.,][]{Kumar-Misra2016a}. 
However, to explain the high energy power-law tail by thermal Comptonization,
the required electron medium temperature is $\sim$100 keV and the optical depth
around
unity \cite[][]{Titarchuk-etal2014,Parker-etal2016}.
Mainly, three models are proposed for power-law tail in context of
Comptonization, i) non-thermal Comptonization \cite[][]{Done-Kubota2006,Kubota-Done2016}
ii) hybrid Comptonization model \cite[][]{Coppi-1999, Gierlinski-etal1999}
iii) bulk Comptonization model \cite[][]{Titarchuk-etal1997, Paizis-etal2006,
Farinelli-etal2009}.
In bulk Comptonization, the soft photons are upscattered by high energetic
electrons, which have both thermal and bulk motions.
For bulk region, \citet{Titarchuk-etal1997} had considered a free-fall
converging flow of spherically accreted plasma into BH (later, a bulk
Comptonization model was also developed for NS with a similar
free-fall bulk region, \citealp[see, e.g.,][]{Farinelli-etal2009}).
However, a spherically free-fall region into BH was first proposed by
\citet{Chakrabarti-Titarchuk1995} to examine the HS state in BH XRBs.

An outflow is observed in both HS and SPL state in many source. In SPL state 
occasionally, an episodic relativistic jet outflow is observed. 
In HS state, a wind outflow is occurred in wide range of speed 0.01--0.03c
and the wind launching radius decreases with increasing wind speed
\cite[e.g.,][]{Trigo-Boirin2016, Tombesi-etal2012,Tombesi-etal2015,Miller-etal2016, Ponti-etal2012}. 
In this work, our primarily motivation is to investigate the outflow geometries
which can generate the high energy power-law tail. As it was earlier
noticed that for spherically divergent type of
outflow, the soft photons spectrum would be got only downscattered \cite[see, e.g.,][]{Psaltis-2001,Laurent-Titarchuk2007,Titarchuk-etal2012, Ghosh-etal2010}.
We use a Monte Carlo scheme for computing the bulk Comptonized spectra.
We find that when the bulk region is due to a conical outflow (of speed greater
than 0.4c) then the bulk Comptonized spectra have power-law component even for
the low Comptonizing medium temperature. Although, such high speed wind ($>$ 0.4c) is not yet observed.
 


\section{Methodology $\&$ Emergent spectrum from outflow geometry}
In bulk Comptonization process the average energy exchange per scattering for a
monochromatic photon
of energy E  \cite[][]{Blandford-Payne1981a, Blandford-Payne1981b,Titarchuk-etal1997} 
is $\Delta$E = $\Delta$E$_{TH}$+ $\left( \frac{4u_b}{c\tau} + \frac{(u_b/c)^2}{3} \right)$ $\frac{E}{m_ec^2}$. Here, $\Delta$E$_{TH}$=  (4kT$_{e}$ - E)
$\frac{E}{m_ec^2}$ is for thermal Comptonization i.e.,
u$_{b}$ = 0, kT$_{e}$ is the Comptonizing medium temperature, u$_{b}$ is the bulk speed, $\tau$ is the optical depth of the scattering medium, m$_e$ is the rest mass of the electron and k is
Boltzmann constant.
For a sets of parameters, if the corresponding
$\Delta$E is same and the bulk direction is taken to be random (like the
thermal motion) then the emergent spectra are similar for a given average scattering
number $\langle N_{sc} \rangle$.

\begin{figure}
\includegraphics[width=0.5\textwidth]{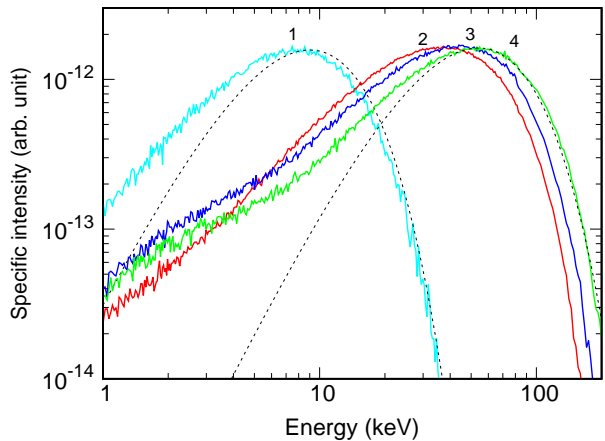} 
\caption{The emergent Wien peak spectrum, the solid lines are for bulk 
Comptonized
and the dashed lines are for thermal Comptonized 
($\propto E^3 exp(-E/kT_e)$) spectra. 
For solid curve 1 the parametrs are (kT$_e$= 2.0keV, u$_b$ = 0.0766c) and for
the dashed curve 1 (kT$_e$= 3.0keV, u$_b$ = 0.0).
The solid curve 2, 3, 4 are for u$_b$ = 0.3c and the kT$_e$ = 0.05, 2.0
and 4 keV respectively, and the correspondingly thermal Comptonized Wien peak
temperatures kT$_e$= $\sim$ 13, 15 and 18 (only shown for curve 4) keV
respectively.
Here, we assume a random bulk motion.}
\label{chk-delE1}
\end{figure}

We compute the bulk Comptonized spectra by Monte Carlo (MC) methods with
neglecting the general relativistic effects, and the algorithm for
MC scheme is 
similar to \citet{Kumar-Misra2016,Laurent-Titarchuk1999, Niedzwiecki-Zdziarski2006}.
Due to the bulk motion of the medium, the mean free path for photon is
increased, e.g., in Table \ref{lambda}, we have listed the variation of mean free path of photon with u$_b$. 
We check the MC code results, by comparing the simulated bulk Comptonized
spectra with thermal
Comptonized spectra (u$_b$ =0). 
For example, for a bulk Comptonization parameter set
(kT$_e$ = 2.0 and u$_b$ = 0.076 c) and a thermal comptonization parameter set
(kT$_e$ = 3.0 keV and u$_b$ = 0), $\Delta$E is same, and we find the
emergent bulk comptonized spectra are similar to the thermal comptonized one
either in case of single scattering or multiple scattering
\cite[see, for details,][]{Kumar2017}. We also
simulate the Wien peak spectra (i.e., $\langle N_{sc} \rangle$ $>$
500) for these two sets and we find the corresponding Wien peak is
similar, as shown by curve 1 in Fig.
\ref{chk-delE1}. Here, One can loosely say that the equivalent thermal
temperature
for bulk parameter set (kT$_e$ = 2.0 and u$_b$ = 0.076 c) is 3.0 keV.
In similar way, we assign the equivalent thermal temperature for bulk
parameters kT$_e$ $\&$ u$_b$ (0.05 keV, 0.3c), (1.0 keV, 0.3c) and
(2.0 keV, 0.3c) which are 14, 16, and 18 keV respectively (as shown in Fig. \ref{chk-delE1}).
We also confirm the
consistency of MC results by comparing the existing results of
\citet{Laurent-Titarchuk2007} for a flat geometry. 

\begin{table}
\tabularfont
\caption{The mean free path of photon of energy E at given bulk speed u$_b$,
when the medium has
temperature kT$_e$ = 3.0 keV and the optical depth $\tau$ = 3. 
}
\label{lambda} 
\begin{tabular} {p{1.15cm}|p{.75cm}p{.750cm}p{.750cm}p{.750cm}p{.750cm}p{.750cm}}
\midline
  E (in keV) & \multicolumn{6}{p{6.5cm}}{the mean free paths of photons (in unit of medium width L) when the bulk speed u$_b$ =}\\
\cline{2-7}
   &  0 & 0.1c & 0.5c & 0.9c & 0.95c & 0.99c  \\ \midline
0.5 & 0.334  & 0.337 & 0.407  & 1.09 & 1.82 & 6.82 \\
1.0 & 0.334 & 0.338 & 0.408 & 1.09 & 1.82 & 6.85 \\
10.0 & 0.346 & 0.350 & 0.423 & 1.14 & 1.91 & 7.23 \\
50.0 & 0.395 & 0.399 & 0.484 & 1.32 & 2.22 & 8.38 \\
100.0 & 0.451 & 0.455 & 0.551 & 1.49 & 2.51 & 9.35 \\
\hline
\end{tabular}
\end{table}

\begin{figure}\vspace{-1.cm}
\centering
\includegraphics[width=0.5\textwidth]{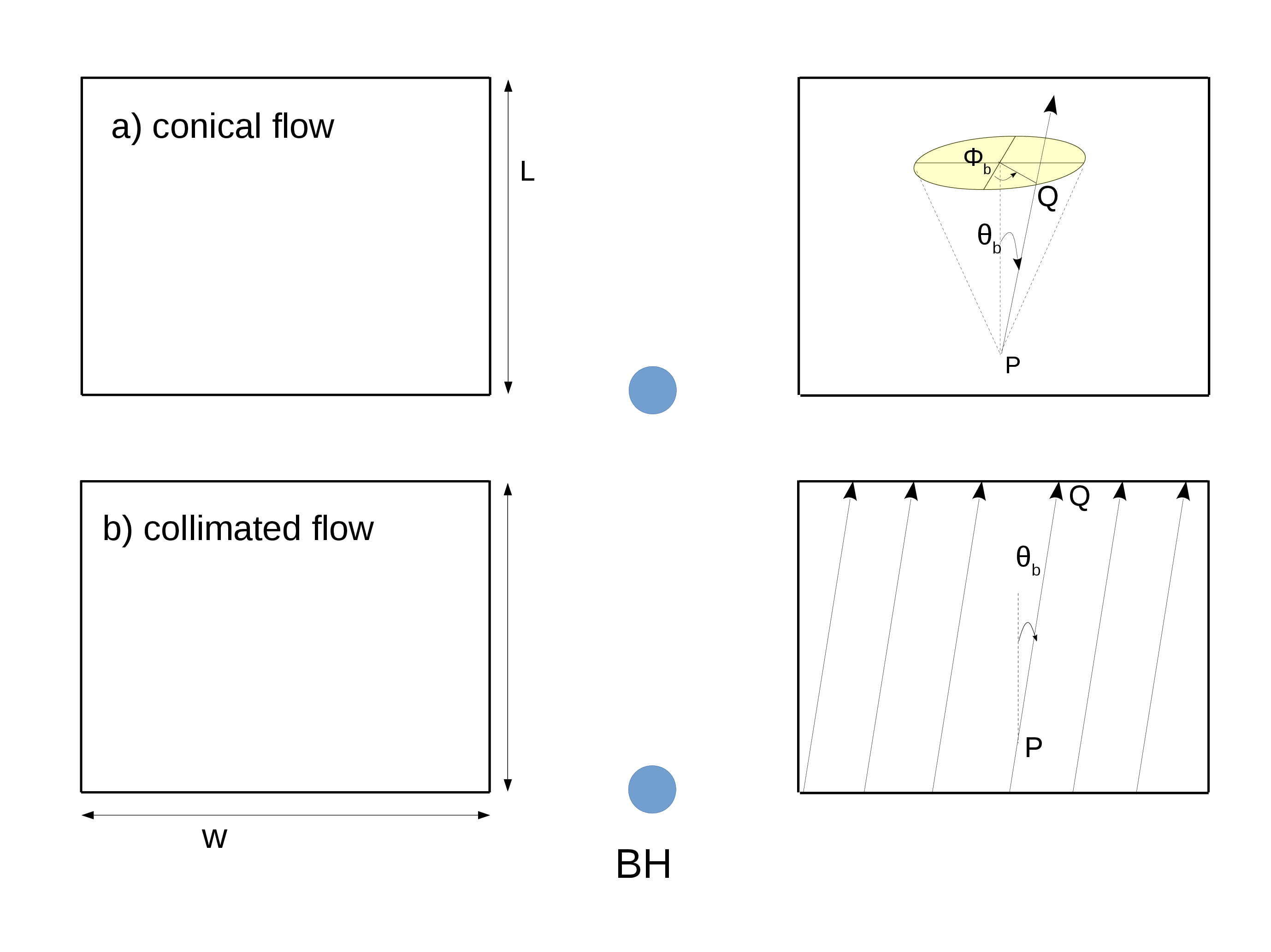} \vspace{-1.cm}
\caption{Meridional cut of a rectangular torus surrounding the compact object
for studying an outflow motion.   
The upper panel is for a conical flow of
opening angle $\theta_b$ and axis is perpendicular to the equatorial plane.
The shaded right circular conic region at scattering point P is 
for the probable bulk direction from the vertex of the cone P.
The lower panel is for collimated flow, where the $\theta$-angle
of bulk direction ($\theta_b$) is same anywhere and the bulk direction is
away from the BH.  
Here, w is width of the torus
and L is the height of the torus but both are not in a scale, and PQ is
a bulk direction, which has $\theta_b$ $\&$ $\phi_b$ angle in local
coordinate.
}
\label{geo-outflow1}
\end{figure}

\begin{figure*}
\centering$
\begin{tabular}{lcr}\hspace{-0.9cm} 
  \includegraphics[width=0.36\textwidth]{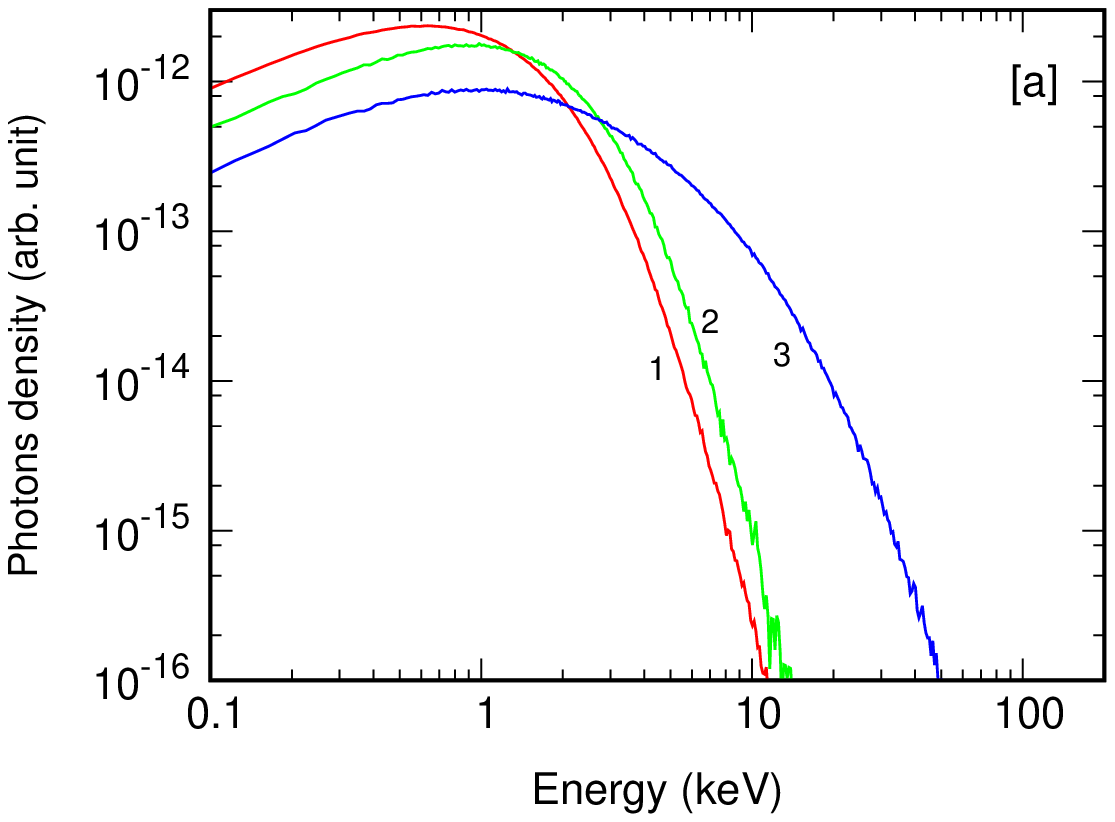} &\hspace{-0.8cm}
  \includegraphics[width=0.36\textwidth]{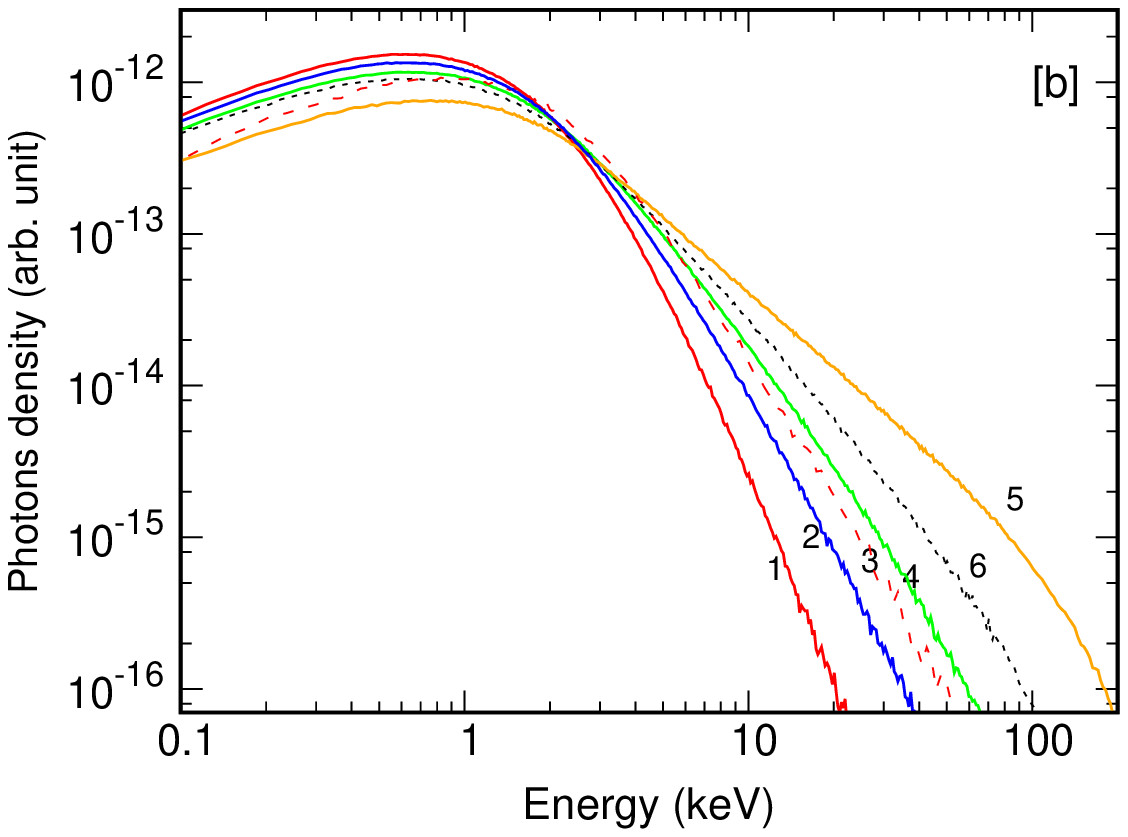} &\hspace{-0.8cm}
  \includegraphics[width=0.36\textwidth]{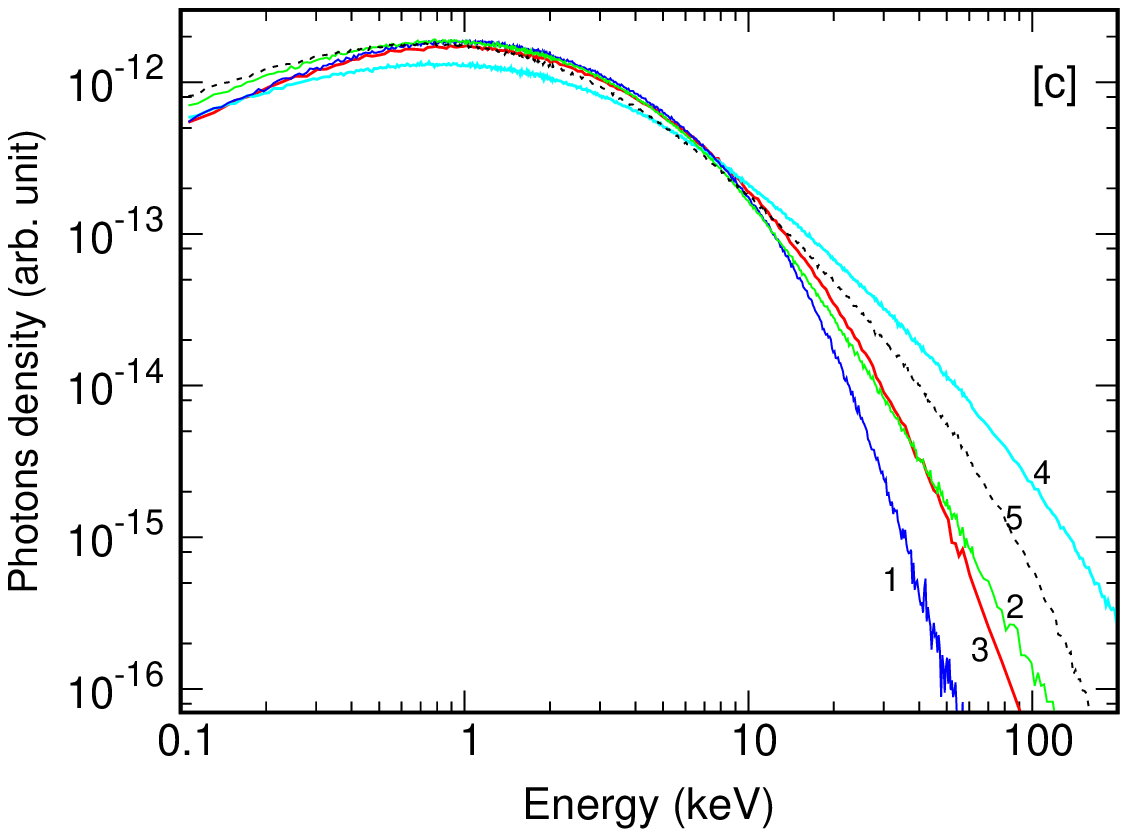} \\
\end{tabular}$
\caption{Emergent bulk Comptonized spectra for different outflows geometry.
In the left panel, the curves 1, 2 and 3 are for $\theta_b$ =0 (outflow),
180 (inflow) and 90 degree respectively in collimated flow. The parameters for
curves 1 and 2 are $\tau$ =3 and u$_b$ =0.45 c, and for curve 3 
$\tau$ =9 and u$_b$ =0.85 c.
The middle panel is for conical flow. In which, the curves 1, 2, 3, 4 and 5 are
for case I with $\theta_b$= 20, 30, 140, 40
and 90 degree respectively and curve 6 is for case II with $\theta_b$=90 degree.
The parameters are  $\tau$ =3 and u$_b$ =0.45 c.
The right panel is for torus flow. The curves 1, 3 are for $\theta_b'$ = 15 degree, and for 2 and
4 
curves $\theta_b'$ is 45 degree. The parameters ($\tau$, u$_b$) are (3, 0.85c) for the
curves 1 and 2; (9, 0.85c) for the curves 3 and 4; and for curve 5, it is 
(9, 0.75c). The rest parameters are kT$_e$ = 3.0 keV, kT$_b$ = 0.5 keV.
}
\label{spec-outflow}
\end{figure*}

We consider a rectangular torus surrounding the compact object
for studying the outflow motion.
We assume, without loss of the generality, the torus exists above the equatorial
plane of the disk, and the scattered photons are absorbed by the disk whenever
it crosses the disk. The torus has width w (=10 km) and vertical height L (= 1
km), we consider the optical depth is in vertical direction, hence the electron
density n$_e$ is $\frac{\tau}{L \sigma_T}$, where $\sigma_T$ is Thompson
cross-section. We assume the soft photon emits vertically from the
equatorial plane of the disk, which is a blackbody at temperature kT$_b$.
We fix a spherical polar global coordinate 
(r, $\theta$, $\phi$) at the compact object and 
we assign the bulk direction, locally, by ($\theta_b$, $\phi_b$) in
same global coordinate.

We consider two different possible outflow
geometries, which originate at the equatorial plane of the accretion disk
as shown in Fig.\ref{geo-outflow1}. a) a right circular conical outflow of
opening angle $\theta_b$ and the axis is perpendicular to the disk
b) a collimated flow with angle $\theta_b$ 
and in local coordinate the bulk direction at any scattering point is ($\theta_b$=constant, $\phi_b$=$\phi_p$), where $\phi_p$ is $\phi$-angle of the scattering
point P in global coordinate.
We also consider a third possibility of a bulk motion, which is an essentially
a collimated flow with $\theta_b$ = 90 degree, and we take $\phi_b$ in such a
way, the bulk velocity corresponds to the azimuthal velocity of the torus fluids, 
so here the materials are not in the outflow but it rotates around the compact
object and we termed as a torus flow.
Further, we assume that the bulk directions vary vertically, i.e., 
$\theta_b$ varies from 90 to
90+$\theta_b'$ degree while  $\phi_b$ fixes to $\phi_p$+90 degree.

The simulated bulk Comptonized spectra due to an outflow from the disk are shown
in Fig. \ref{spec-outflow}.
The emergent spectra for collimated flow has been shown in Fig. \ref{spec-outflow}a. 
As expected, in case of curve 1 ($\theta_b$ = 0 degree, outflow from the disk), the
photons get downscattered 
since these photons are hitting outflows from behind. But in case of curve 2
($\theta_b$ = 180 degree, inflow towards the disk) the photon and
electron experience a head-on collisions in first scattering, however after
few scattering, photon tends to move in electron direction due to the
high speed
of the electron, mainly in electron's bulk direction (due to bulk dominated case, i.e., $(u_b/c)^2$ $\gg$ $3kT_e/(m_ec^2)$), so again the photons are hitting outflows from
behind and it gets downscattered.  \citet{Janiuk-etal2000} had
also found the similar results, a multiple scattered inflow spectra would be
softer in comparison to the single scattered spectra. 
We find that there are no high energies power-law tail in spectrum 
for given kT$_e$ = 3.0 keV, even for the extreme value of parameters, like
u$_b$= 0.85c and $\tau$= 9, as shown by the curves 3.

The emergent spectra due to a conical flow have been presented in Fig. \ref{spec-outflow}b. We consider two different cases for bulk direction in conical flow,
in case I, the bulk directions are in any one of direction along the surface of
the cone i.e., $\theta_b$ = constant and $\phi_b$ varies from 0 to 360 degree.
In case II, the bulk directions are in any one of direction inside the conical
region i.e., $\theta_b$ varies from 0 to $\theta_b$, and $\phi_b$ varies
as the case I.
The curves 1, 2, 3, 4 and 5 are for $\theta_b$ = 20, 30, 140, 40 and 90 degree
respectively for case I and the curve 6 is for case II with
$\theta_b$ = 90 degree. Here, we like to mention that $\theta_b$ = 90 degree
is not looks like a conical flow but for completeness we present the spectra. 
We observe that the spectrum is similar for the conical outflow and inflow with
having same opening angle, which is shown by curves 3 and 4 in Fig.
\ref{spec-outflow}b. 
In both cases, we find that the photon index decreases with $\theta_b$ and
correspondingly the high
energy cutoff ($E_{cut}$) of the spectra increases, and the
spectra of case I are harder than case II.
For example, in case I, $\Gamma$ varies from $\sim$4. to $\sim$ 1.7 with
varying $\theta_b$ 20 to 90 degree respectively and corresponding $E_{cut}$
changes 20 to 200 keV (as shown in Fig. \ref{spec-outflow}b).
Hence, to generate the observed
high energy power-law tail, i.e., $\Gamma$ $>$ 2.4 and $E_{cut}$ $>$ 200 keV,
the bulk speed must be greater than 0.4c. 

Unlike the collimated flow, the high energies power-law tail can be
generated 
in a conical flow. As mentioned, for a collimated flow the bulk
directions are fixed in each event (here, the event means a track of a photon
inside the scattering medium),
while in case of conical flow the bulk directions would be 
varied due to the variations of $\phi_b$. So the angle between
photons and electrons will vary from 0 to $\theta_b$ in each event.
Hence the probability for getting
a behind hit of outflow by photon decreases with  $\theta_b$.
Thence after certain lower value of $\theta_b$ photons get upscattered, and
which is around 30 degree.
We notice that the high energies power-law tail (of observed $\Gamma$ value)
can be extended upto
200 keV only when $\theta_b$ greater than 30 degree and u$_b$ $>$ 0.4c.
\cite[see, for details,][]{Kumar2017}.

In Fig. \ref{spec-outflow}c, we presented the emergent bulk Comptonized
spectra due to
the azimuthal velocity of the torus fluid.
We compute this spectra for two
values of $\theta_b'$= 15 and 45 degree. The photon index of the power-law
tail decreases with increasing $\theta_b'$, and the increasing
$\theta_b'$ means
the bulk direction becomes more random, for example $\theta_b'$ = 15
degree, the 
$\theta_b$ varies from 90 to 105 degree. 
It seems that to obtain the power-law tail, which is extended at least
upto 200 keV, the $\theta_b'$ should be greater than 15 degree (see, curve 3 
in Fig. \ref{spec-outflow}c), u$_b$ should be larger than 0.75c (see, curve 5 in Fig. \ref{spec-outflow}c). 
Hence, the conical outflow with opening angle greater than $\sim$30 degree, and 
the rotating plasma around the compact objects with high speed (u$_b$ $>$ 
0.75c, $\&$ $\theta_b'$ $>$ 15) and large optical depth are a possible bulk region which can generate
the power-law tail with observed range of $\Gamma$ by bulk Comptonization
process even at low medium temperature, while the collimated flow is not a
plausible bulk 
region to generate the power-law tail at low medium temperature.  

\section{Summary and Discussion}
The high energy power-law tail of HS state and the SPL state of BH XRBs 
can be generated by a bulk Comptonization process with having spherically 
free-fall bulk region into the compact object \cite[see, e.g.,][]{Laurent-Titarchuk1999, Laurent-Titarchuk2011}.
Although, for a spherically divergent flow, these power-law tail can not be
produced \cite[e.g.,][]{Psaltis-2001,Laurent-Titarchuk2007,Ghosh-etal2010}. 
However \citet{Titarchuk-etal2012} had explained the power-law components of gamma-ray burst
spectrum by the bulk Comptonization due to a subrelativistic outflow, but they
considered a thermal motion dominated regime ($(u_b/c)^2$ $\ll$ $3kT_e/(m_ec^2)$).
Moreover, \citet{Gierlinski-etal1999} emphasized that a bulk Comptonized spectra due to a free-fall
bulk region can not be extended more than 200 keV \cite[see, also,][]{Revnivtsev-etal2014, Zdziarski-etal2001}.
In this work, we generate the high energy power-law tail in bulk Comptonized
spectra by using a Monte Carlo scheme with considering an outflow from
the disk as a bulk
region, and these spectra can be extended more than 200 keV depending upon
(mainly) the outflow speed.

For an outflow region, we consider a rectangular torus surrounding
the compact object, and we assume that the soft photon source is inside the
torus which emits vertically from the equatorial plane of the disk.
We first consider 
a torus flow, where the fluid rotates around
the compact object and its azimuthal velocity serves as a bulk motion, and so
it is not an outflow type of geometry. \citet{Abramowicz-etal1978} shown that
an equilibrium perfect fluid torus can be presented at inner edge of the
accretion flow \cite[see, also,][]{Abramowicz-etal2006}, and
the epicyclic mode of the torus can be a plausible mechanism to produce 
a observed high frequency QPOs ($>$ 100 Hz) of blackhole X-ray
binaries.
Since the high frequency QPOs usually observe in SPL state, so we motivate
to study with this question that for what situations the torus flow can
generate the high energy power-law tail by bulk Comptonization process.
We find, the power-law tail
can generate in case of the torus flow with high optical depth and large bulk speed, but here we assume
an ad hoc variation in bulk direction, i.e., only in vertical direction,
which may be happened due to the epicyclic mode, a consequence of the
perturbation of the  
gravity of the compact object. But it should be
incorporated in self-consistent manner, we intended to do in subsequent
paper with general relativistic effect.

Next we consider two other outflow geometry, one is a collimated flow and other
one is a right circular conical flow (of opening angle $\theta_b$ and the
axis is perpendicular to the disc). For a conical flow,
two different bulk directions have been considered, in case I the bulk
direction can be any one of direction along the surface of the cone, and in case
II the bulk direction can be any one of direction inside the conical region.
We find that the spectra for case I is harder than case II, and in both cases
$\Gamma$ decreases with $\theta_b$.
For a collimated flow, the high energy power-law tail can not be generated,
but in case of a conical flow, the power-law tail can be generated
with opening angle
$\theta_b$ larger than $\sim$30 degree for a low Comptonizing
medium temperature.
We notice that in conical flow to generate the power-law tail
for given photon index, the
lower $\theta_b$ has larger $\tau$ at given u$_b$ and it has high value of
u$_b$ at given $\tau$ in comparison to higher $\theta_b$.
The lower opening angle conical outflow is a plausible
bulk region to produce the high energy power-law tail in both the spectral states HS
and SPL for $E_{cut}$ $<$ 1000 keV \cite[see, for details,][]{Kumar2017}. But, the bulk speed should be
larger than 0.4c, and such high speed wind have not observed yet. If the high
speed wind is occurred in X-ray emitting region, then this study may serve to
study the outflow properties. Although, the high speed outflow is observed in
jet flow, but it is generally believe that jet is a collimated flow not a
conical flow. But, if the SPL spectra characterize by low medium temperature,
it may have an implication to
understand the some observed features of gamma-ray burst spectrum, or
broad band blazar spectrum \cite[e.g.,][]{Kushwaha-etal2014}. 
However, we do not compare the emergent spectrum with observed spectrum,
which we intended to do in a future work, also it is needed to find the physical
mechanism for such a conical
bulk flow which will constrain its physical parameters
like bulk speed, the angular distribution and also the optical depth.

\section*{ACKNOWLEDGEMENTS}
NK acknowledges financial support from Indian Space Research Organisation (ISRO)
with research Grant No.
ISTC/PPH/BMP/0362. NK wishes to thank Ranjeev Misra for valuable comments on
this project and Banibrata Mukhopadhyay for their valuable
suggestions and comments over the manuscript.

\def\aap{A\&A}%
\def\aapr{A\&A~Rev.}%
\def\aaps{A\&AS}%
\def\aj{AJ}%
\def\actaa{Acta Astron.}%
\def\araa{ARA\&A}%
\def\apj{ApJ}%
\def\apjl{ApJ}%
\def\apjs{ApJS}%
\def\apspr{Astrophys.~Space~Phys.~Res.}%
\def\ao{Appl.~Opt.}%
\def\aplett{Astrophys.~Lett.}%
\def\apss{Ap\&SS}%
\def\azh{AZh}%
\def\bain{Bull.~Astron.~Inst.~Netherlands}%
\def\baas{BAAS}%
\def\bac{Bull. astr. Inst. Czechosl.}%
\def\caa{Chinese Astron. Astrophys.}%
\def\cjaa{Chinese J. Astron. Astrophys.}%
\def\fcp{Fund.~Cosmic~Phys.}%
\def\gafd{Geophys.\ Astrophys.\ Fluid Dyn.}
\def\gca{Geochim.~Cosmochim.~Acta}%
\def\grl{Geophys.~Res.~Lett.}%
\def\iaucirc{IAU~Circ.}%
\def\icarus{Icarus}%
\def\jcap{J. Cosmology Astropart. Phys.}%
\def\jcp{J.~Chem.~Phys.}%
\def\jfm{JFM}
\def\jgr{J.~Geophys.~Res.}%
\def\jqsrt{J.~Quant.~Spec.~Radiat.~Transf.}%
\def\jrasc{JRASC}%
\def\mnras{MNRAS}%
\def\memras{MmRAS}%
\def\memsai{Mem.~Soc.~Astron.~Italiana}%
\def\na{New A}%
\def\nar{New A Rev.}%
\def\nat{Nature}%
\def\nphysa{Nucl.~Phys.~A}%
\def\pasa{PASA}%
\def\pasj{PASJ}%
\def\pasp{PASP}%
\def\physrep{Phys.~Rep.}%
\def\physscr{Phys.~Scr}%
\def\planss{Planet.~Space~Sci.}%
\def\pra{Phys.~Rev.~A}%
\def\prb{Phys.~Rev.~B}%
\def\prc{Phys.~Rev.~C}%
\def\prd{Phys.~Rev.~D}%
\def\pre{Phys.~Rev.~E}%
\def\prl{Phys.~Rev.~Lett.}%
\def\procspie{Proc.~SPIE}%
\def\qjras{QJRAS}%
\def\rmxaa{Rev. Mexicana Astron. Astrofis.}%
\def\sgg{Stud.\ Geoph.\ et\ Geod.}
\def\skytel{S\&T}%
\def\solphys{Sol.~Phys.}%
\def\sovast{Soviet~Ast.}%
\def\ssr{Space~Sci.~Rev.}%
\def\zap{ZAp}%
\def\memsai{Memorie della Societa Astronomica Italiana}


\end{document}